\documentclass{iau} 
\usepackage{graphicx}

\newcommand{\her}{{\it Herschel}}

\newcommand{\spitz}{{\it  Spitzer}}
\newcommand{\iso}{{\it ISO}}
\newcommand{\iras}{{\it IRAS}}

\newcommand{\akari}{{\it AKARI}}

\newcommand{\zsun}{$Z_\odot$}
\newcommand{\mic}{$\mu$m}

   \newcommand{\hmol}{H$_2$}

 \newcommand{\ciifir}{L$_{[C\,{\sc II}]}$/L$_{FIR}$}
  
 \newcommand{\ciico}{L$_{[C\,{\sc II}]}$/L$_{CO(1-0)}$}

\title[IAU 315~~Low Metallicity Dwarf Galaxies] 
{How Does Metallicity Affect  the Gas and Dust Properties of Galaxies?}

\author[Suzanne C. Madden \& Diane Cormier \& Aurelie Remy-Ruyer]    
{Suzanne C. Madden$^1$,
Diane Cormier$^2$ \and Aur\'elie R\'emy-Ruyer$^{3,1}$}

\affiliation{$^1$CEA/DSM/IRFU/Service d'Astrophysique, AIM (UMR 7158),
91191, Gif-sur-Yvette, France \\ email: {\tt suzanne.madden@cea.fr} \\[\affilskip]
$^2$Institute for Theoretical Astrophysics, Center for Astronomy, University of Heidelberg, \\ 69120, Heidelberg, Germany \\email: {\tt diane.cormier@zah.uni-heidelberg.de} \\[\affilskip]
$^3$Institut d'Astrophysique Spatiale, CNRS, 91405, Orsay, France \\email: {\tt ruyeraurelie.astro@gmail.com}}

\pubyear{2015}
\volume{315}  
\jname{From Interstellar Clouds to Star-Forming Galaxies: Universal Processes ?}
\editors{P. Jablonka, P. Andre \& F. van der Tak, eds.}
\begin{document}

\maketitle

\begin{abstract}
 Comparison of the ISM properties of a wide range of metal poor galaxies with normal metal-rich galaxies reveals striking differences. We find that the combination of the low dust abundance and the active star formation results in a very porous ISM filled with hard photons, heating the dust in dwarf galaxies to overall higher temperatures than their metal-rich counterparts. This results in photodissociation of molecular clouds to greater depths, leaving relatively large PDR envelopes and difficult-to-detect CO cores. From detailed modeling of the low-metallicity ISM, we find significant fractions of CO-dark H$_{2}$ - a reservoir of molecular gas not traced by CO, but present in the [CII] and [CI]-emitting envelopes. Self-consistent analyses of the neutral and ionized gas diagnostics along with the dust SED is the necessary way forward in uncovering the multiphase structure of galaxies.
  
\end{abstract}

\firstsection 
\section{Introduction}
Dwarf galaxies, often carrying signatures of chemical youth, are unique laboratories to obtain insight on star formation and interstellar medium (ISM) properties in low-metallicity environments that may be relevant to understanding early universe conditions, when metal abundances were very low. The decrease of metals reduces shielding effects which have important consequences, for example, on the molecular gas reservoir, which, in turn, should affect how we observe the ISM properties.  Our knowledge in this subject has been growing at a fast pace due to the increasingly sensitive infrared (IR) space missions \iras, \iso, \spitz, \akari\ and \her, as well as ever-growing powerful ground-based millimeter (mm) and submillimeter (submm) telescopes. The Dwarf Galaxy Survey (DGS; \cite[Madden et al. 2013)]{madden13}, has compiled a large observational database of 48 low-metallicity galaxies, motivated by \her\ PACS \cite[(Poglitsch et al. 2010)]{poglitsch10} and SPIRE \cite[(Griffin et al. 2010)]{griffin10} photometric and spectroscopic observations covering 55 to 500 \mic.  These data, together with the ancillary database, create a rich legacy for dust and gas analyses in unprecedented detail in low-metallicity galaxies from ultraviolet (UV) to mm wavelengths in environments covering the largest metallicity range achievable in the local Universe, as low as $\sim$1/40\zsun \cite[(R\'emy-Ruyer et al. 2013; 2015)]{remy15, remy13}. Here we describe enigmatic issues surrounding the nature of the dust and gas properties in low-metallicity galaxies.

\section{Metallicity effects on dust properties and gas-to-dust mass ratios}
\subsection{Dust spectral energy distributions and dust temperatures}
The dust spectral energy distributions (SEDs) of the DGS galaxies and the \her\ KINGFISH survey, consisting of mostly metal-rich galaxies \cite[(Kennicutt et al. 2011; Dale et al. 2012)]{kennicutt11, dale12} have been modeled by \cite[R\'emy-Ruyer et al. (2015)]{remy15} using the model of \cite[Galliano et al. (2011)]{galliano11}. This complete, consistently modeled sample consists of 109 galaxies combined, covering almost 2 dex in metallicity, making it a valuable sample to investigate the effects of metallicity on the SED shape and the dust properties.
 
\begin{figure}[!ht]
  \begin{center}
\includegraphics*[width=2.55in]{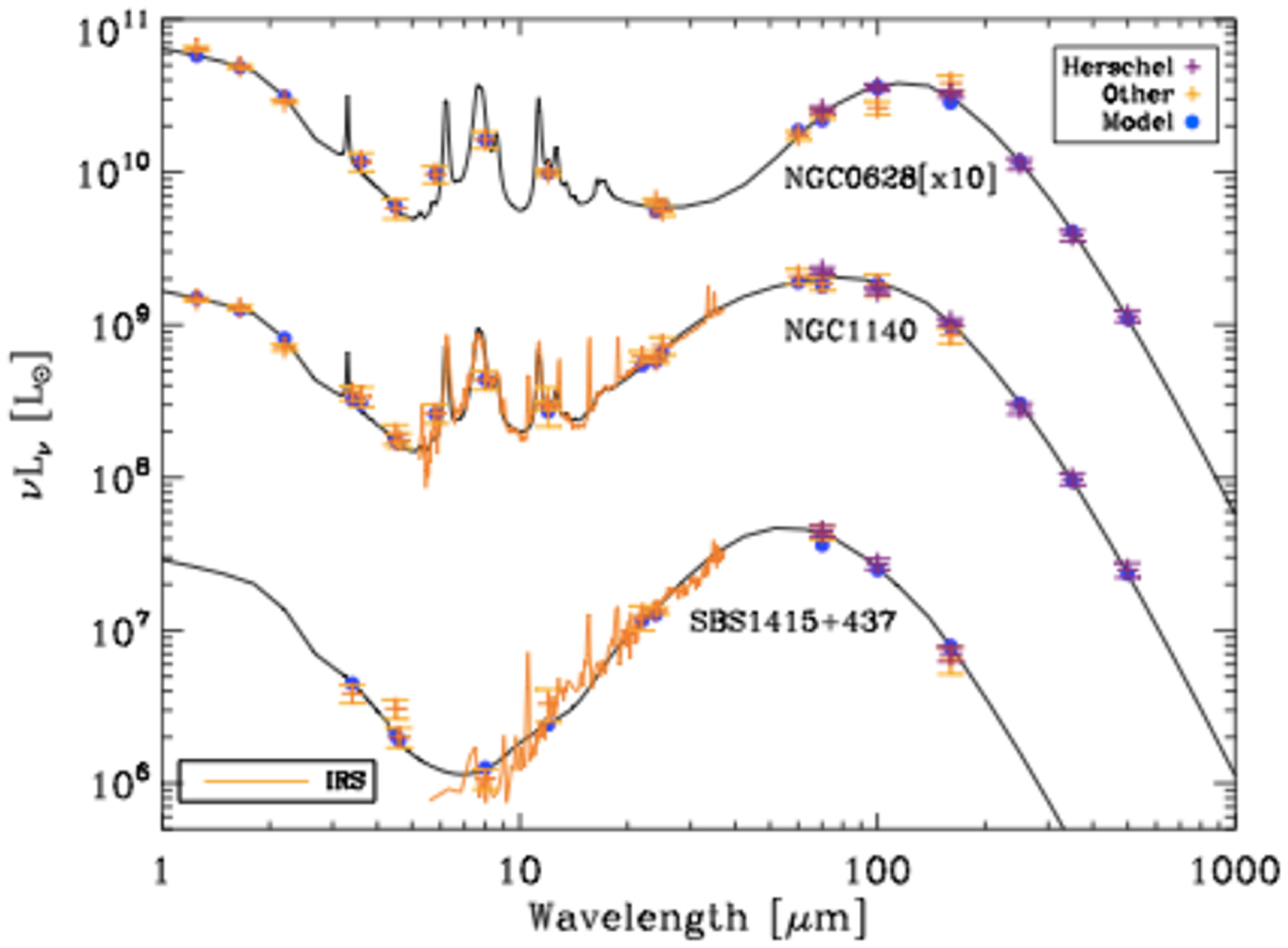}
\includegraphics*[width=2.7in]{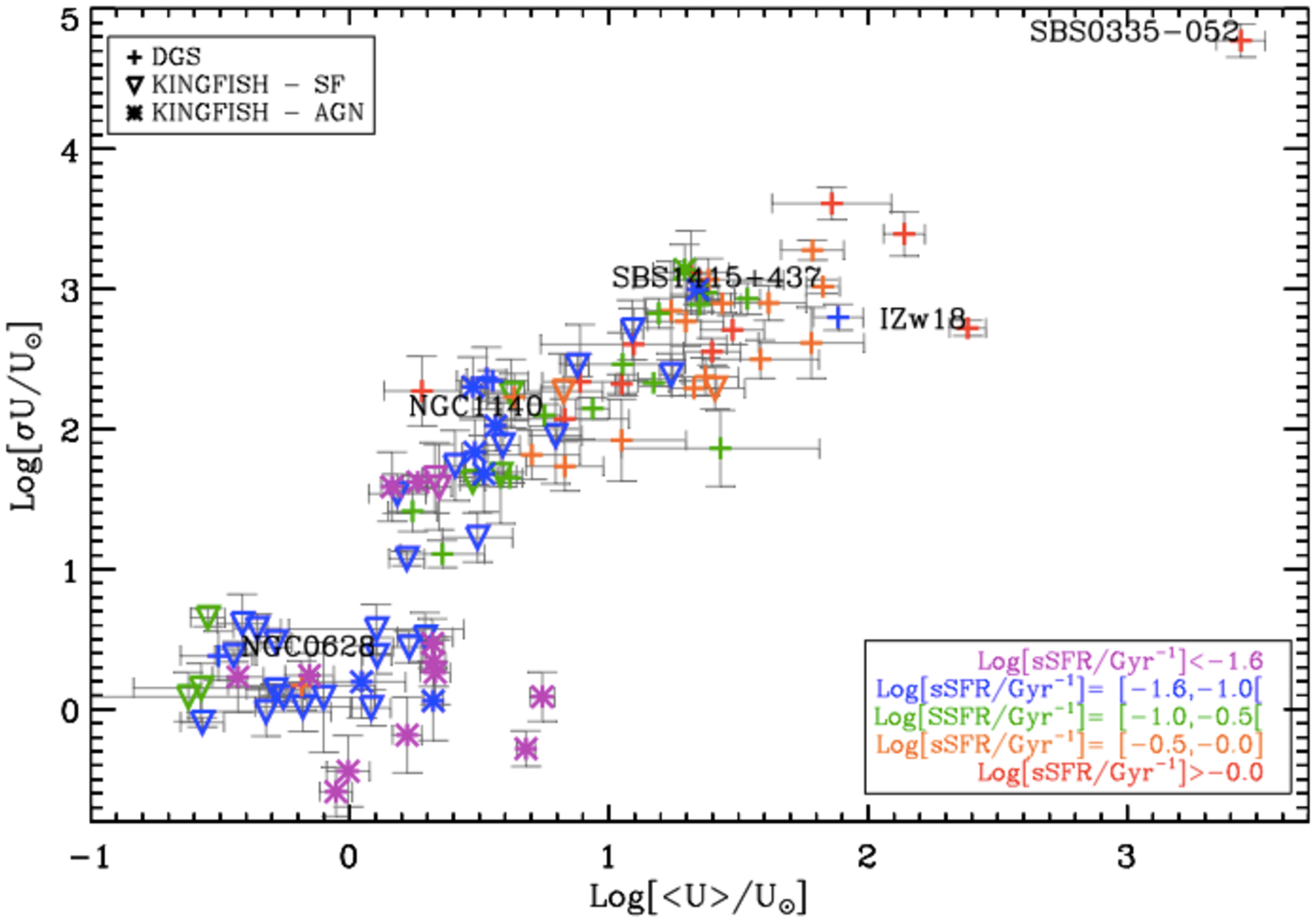}
\caption{\footnotesize{ {\it (left)} Examples of SEDs, demonstrating the increasing broadening of the SEDs, or the wider distribution of the starlight intensities ($\sigma$U). Metallicity is decreasing from the metal-rich galaxy, NGC0628, to the metal-poor galaxy, SBS1415+437. {\it (right)} Broadening of the SED ($\sigma$U) as a function of average starlight intensity ($<U>$) for the DGS + KINGFISH galaxies. 
The color-code is the specific star formation rate (sSFR) of the sources. Notice that the low-metallicity dwarf galaxies tend to have higher $<U>$ and higher $\sigma$U (broader dust SEDs) while the more metal-rich KINGFISH galaxies tend towards lower sSFR values, lower $<U>$ (lower T$_{d}$), and narrower SEDs (see \cite[R\'emy-Ruyer et al. 2015]{remy2015} for more details).}}
\label{seds_sigmaU}
 \end{center}
\end{figure}

Low-metallicity galaxies tend to have higher average starlight intensity ($<U>$) and broader SED peaks ($\sigma$U) shifted to shorter wavelengths compared to the more metal-rich KINGFISH galaxies (Fig.~\ref{seds_sigmaU}, {\it left}), as a result of their high star formation activity. This is reflected in the overall dust temperatures (T$_{d}$): the low-metallicity galaxies span a broader range in temperature and a higher mean T$_{d}$ of 26\,K,  in contrast to the higher-metallicity KINGFISH galaxies which span a narrower temperature distribution and a lower mean T$_{d}$ of 20\,K \cite[(R\'emy-Ruyer et al. 2013; 2015)]{remy13, remy15}. A trend of increasing T$_{d}$ with decreasing metallicity is likewise observed. Also seen in Fig.~\ref{seds_sigmaU}~{\it (right)} is the trend of increasing $<U>$ and $\sigma$U with increasing sSFR showing that the sSFR is the parameter controlling the dust SED shape, while we observe a weaker correlation of $<U>$ and $\sigma$U with metallicity, placing metallicity in a secondary role in shaping the observed SED. The DGS galaxies contain a wider range of warmer dust, primarily due to their intensive star formation activity, not necessarily due to the fact that they are low metallicity. The secondary effect of the low metallicity is reflected in the higher T$_{d}$. The consequence of the lower dust attenuation is the heating of the dust deeper into the molecular clouds.
 
\subsection{Gas-to-dust mass ratio versus metallicity}
\begin{figure}[!ht]
  \begin{center}
 \includegraphics*[width=2.55in]{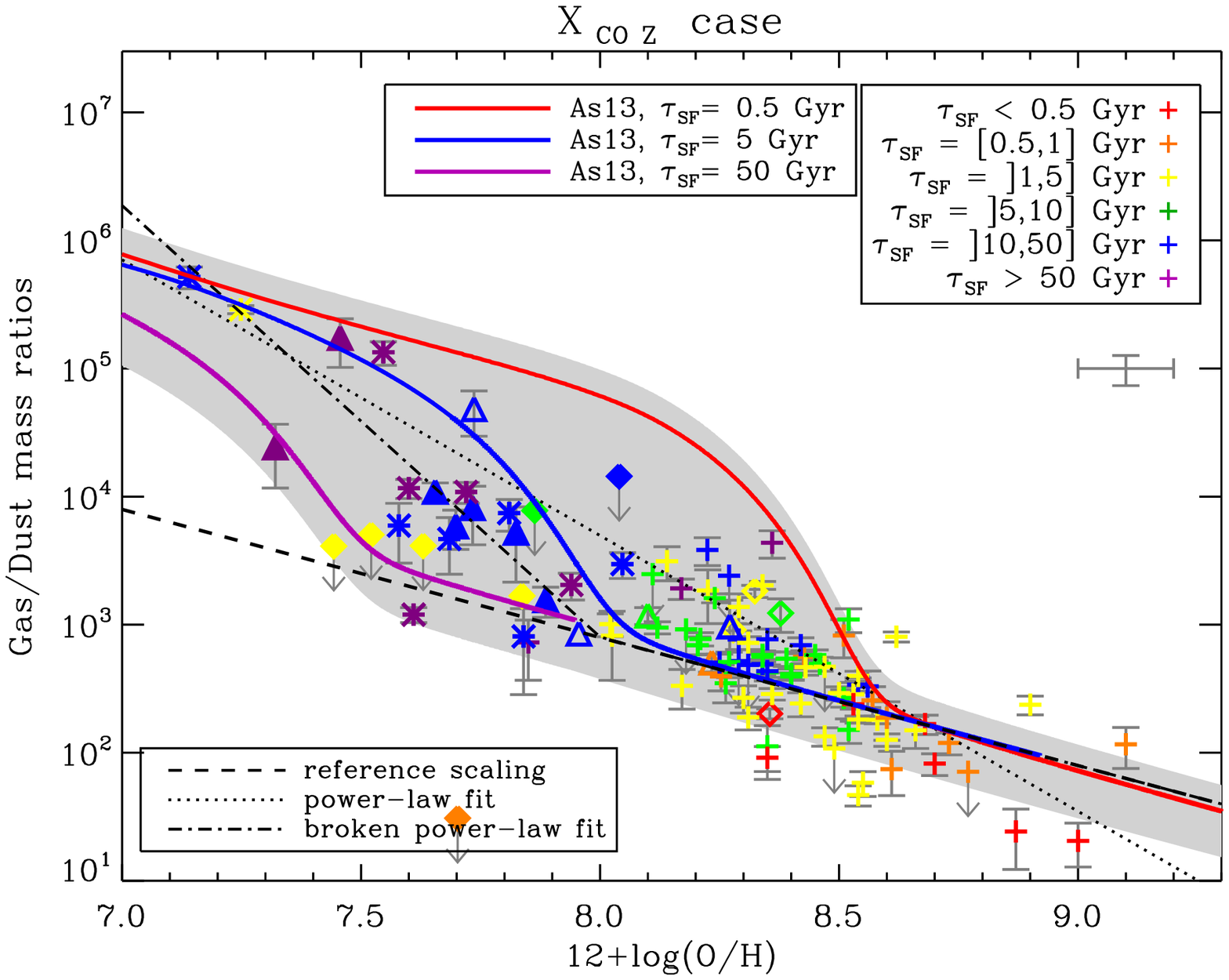} 
 \includegraphics*[width=2.7in]{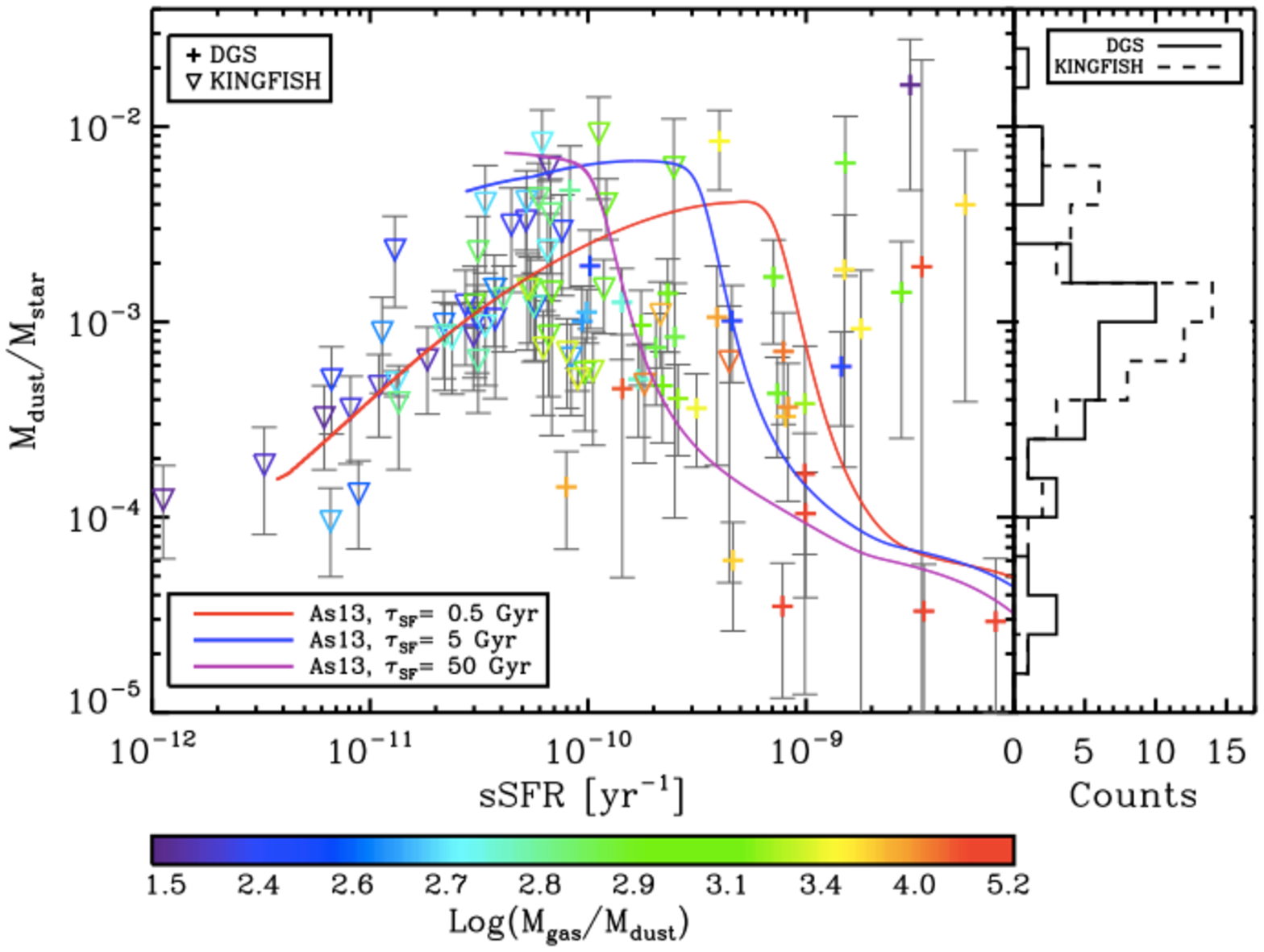}
 \caption{\footnotesize{{\it (left)} Observed gas-to-dust mass ratio (G/D) for the DGS + KINGFISH galaxies compared to models of \cite[Asano et al. (2013)]{asano13} for various star formation time scales ($\tau$) and star formation histories. From the study of \cite[R\'emy-Ruyer et al. (2014)]{remy14}. {\it (right)} Dust-to-stellar mass ratios for the DGS (crosses) and KINGFISH (downward triangles) galaxies as a function of sSFR, color-coded by M$_{gas}$/M$_{dust}$ (from \cite[R\'emy-Ruyer et al. 2015]{remy15}). The distribution of M$_{dust}$/M$_{star}$ is indicated on the side: solid line for DGS and dashed line for KINGFISH.}}
\label{GD_models}
\end{center}
\end{figure}

With dust masses accurately and consistently measured for 109 DGS plus KINGFISH galaxies over the full MIR to submm wavelength range and covering a broad metallicity range, it is now possible to characterize the impact of metallicity on the gas-to-dust mass ratio (G/D) over almost 2 dex range of metallicity (Fig.~\ref{GD_models}, {\it left}; \cite[R\'emy-Ruyer et al. 2014]{remy14}). 
 
We observe a broad degree of scatter over all metallicities and a steep rise in the G/D for the lower-metallicity galaxies for which 12 + log (O/H) $\lesssim$ 8. This behavior is found to be consistent with chemical evolution models that require grain growth in the ISM as a source of dust production as well as episodic star formation history \cite[(e.g. Zhukovska et al. 2014; Asano et al. 2013)]{zhukovska13, asano13}. Dust production becomes more efficient when enough metals can accumulate, and this occurs in the observations and models at around 12 +log (O/H) $\sim$ 8. Normalizing the dust mass by the stellar mass gives us a view of how the dust mass builds itself up as a function of the stellar mass. Previous studies of the dust-to-stellar mass ratio of mostly rather metal-rich galaxies \cite[(e.g. da Cunha et al. 2010; Skibba et al. 2011; Cortese et al. 2012)]{dacunha10, skibba11, cortese12} have shown that the dust-to-stellar mass ratio decreases for increasing stellar mass (or metallicity) and decreasing sSFR.  We extend the dust-to-stellar mass ratio studies to lower-metallicity, low-mass galaxies with our DGS galaxies and we do not find any correlation for the dust-to-stellar mass ratio with metallicity or sSFR over the full sample. The results for only the more metal-rich KINGFISH galaxies are in agreement with the findings of previous studies. However, the low-metallicity sample does not extend the same behavior of the metal-rich galaxies to higher sSFR. Instead we see a wide scatter in dust-to-stellar mass ratio for the DGS galaxies and a very low dust-to-stellar mass ratio for some of the lowest metallicity (high G/D) galaxies (Fig. \ref {GD_models},~{\it right}). The different behavior of the dust-to-stellar mass ratios for the low-metallicity DGS galaxies can be explained by their chemical evolutionary stage, also traced by the G/D \cite[(R\'emy-Ruyer et al. 2014)]{remy14}. The evolutionary tracks of the chemical evolution model of \cite[Asano et al. 2013]{asano13} (Fig. \ref{GD_models}) show a turnover with the scatter related to various star formation time scales ($\tau$) and star formation histories. Dust growth processes saturate when all the available metals are locked up in the grains. In the meantime, star formation continues, consuming the gas reservoir and increasing the stellar mass. This results in increasing the sSFR, decreasing the dust-to-stellar mass ratio and decreasing the G/D. The prominent non-linearity of the G/D vs metallicity at low metallicity, as well as the wide degree of scatter, even for the metal-rich galaxies, call for caution when attempting to determine the total gas mass from the dust mass with an assumed G/D simply scaled linearly with metallicity.

\section{The FIR fine-structure lines reveal a porous ISM structure}
\begin{figure}[!ht]
\includegraphics[width=2.8in]{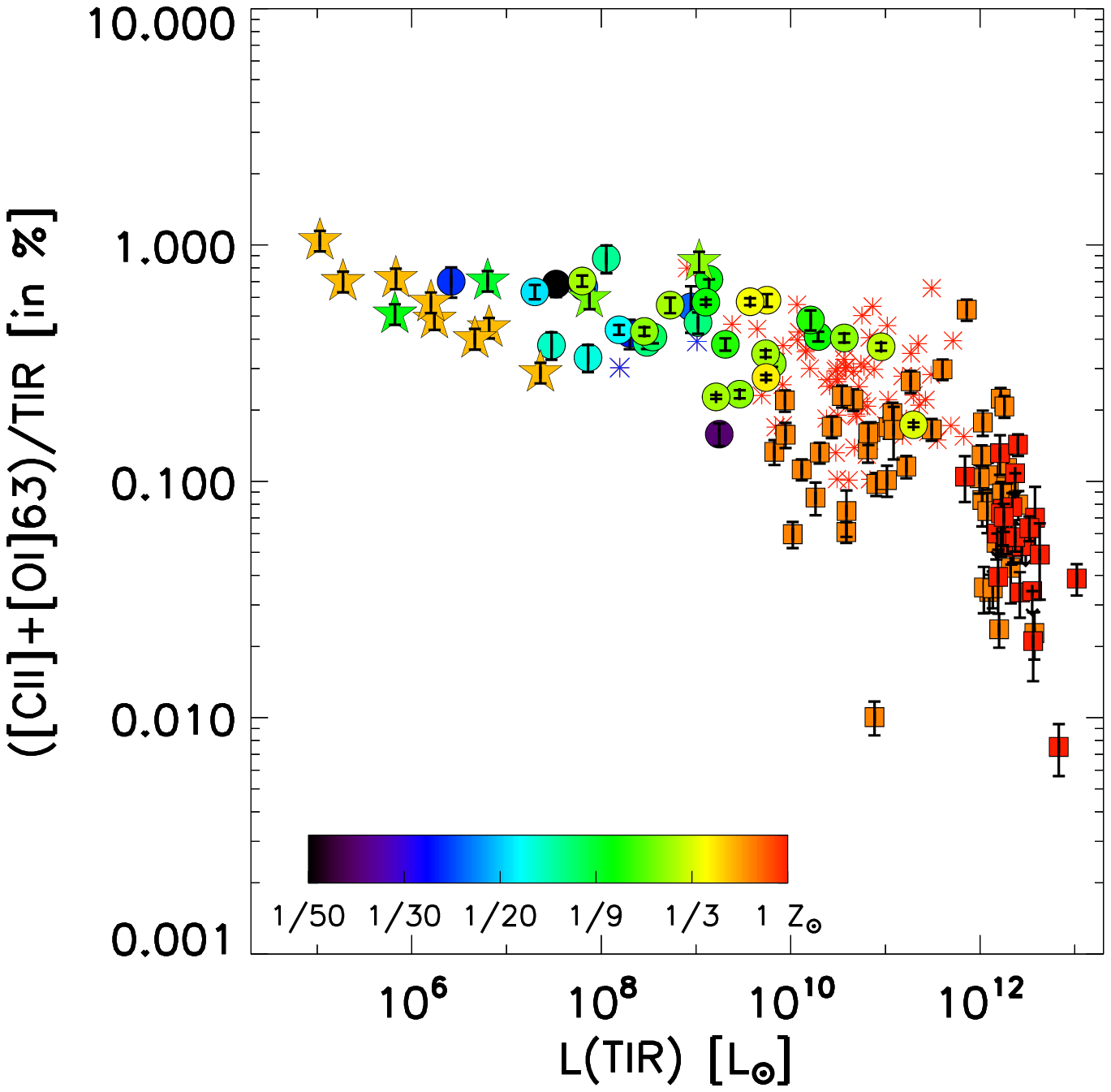}
\includegraphics[width=2.6in]{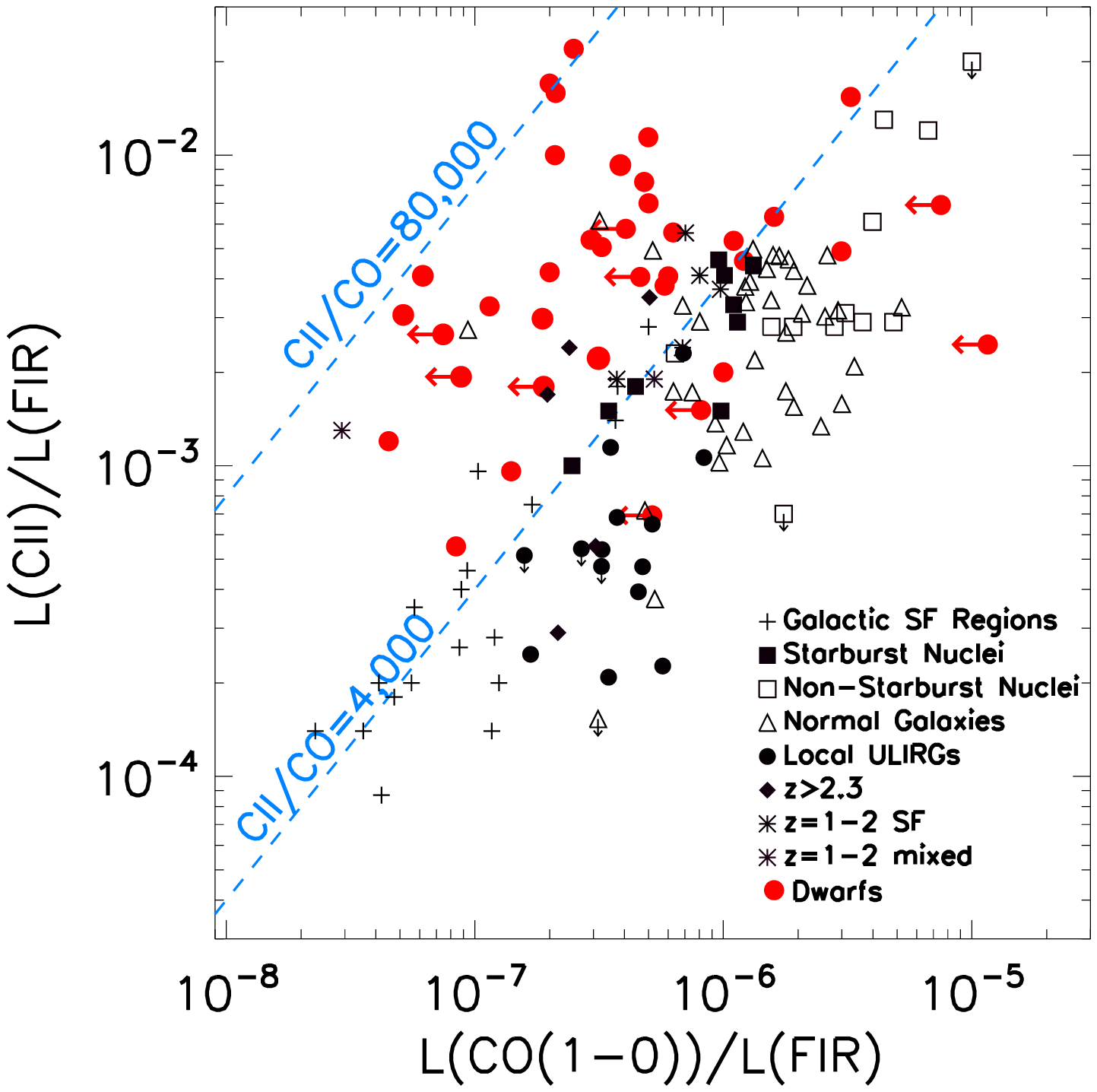}
\caption{\footnotesize{ {\it (left)} ([CII]+[OI])/L$_{TIR}$ vs L$_{TIR}$, from \cite[Cormier et al. (2015)]{cormier15}. {\it (right)} [CII]/L$_{FIR}$ vs CO(1-0)/L$_{FIR}$ for the dwarf galaxies (red symbols) and a wide range of metal-rich galaxies (Madden et al. in prep.). The metal-rich galaxies, black closed and open symbols, are from \cite[Stacey et al. (1991)]{stacey91}.}}
\label{CII_OI_CO}
\end{figure}

The DGS galaxies were surveyed in the most important FIR fine-structure lines, such as 158 \mic\ [CII], 63 and 145 \mic\ [OI], 88 \mic\ [OIII] and more rarely, due to sensitivity limits, 57 \mic\ [NIII], 122 and 205 \mic\ [NII] (see \cite[Madden et al. 2013 and Cormier et al. 2015]{madden13, cormier15} for more details of the DGS observations). These important cooling lines are diagnostics probing the FUV flux, the gas density and temperature, and the filling factor of the ionized gas and photodissociation regions \cite[(PDRs; e.g. Wolfire et al. 1990; Kaufman et al. 2006; Le Petit et al. 2006)]{wolfire90,kaufman06,lepetit06}. The photoelectric effect is normally the dominant source of gas heating in PDRs and [CII] usually ranks foremost in the PDR cooling lines in galaxies, followed by the 63 \mic\ [OI]. Thus the [CII]/L$_{FIR}$ ratio, or ([CII]+ [OI]/L$_{FIR}$, indicates the efficiency of photoelectric heating. Fig.~\ref{CII_OI_CO}~{\it (left)} shows the ([CII]+ [OI])/L$_{FIR}$ ratio as a function of L$_{FIR}$. The ratios of dwarf galaxies are higher than those of galaxies in other surveys of mostly metal-rich galaxies. These relatively larger efficiency factors, from 1\% to 2\%, can be a consequence of relatively normal PDR gas densities (often on the order of 10$^3$ to 10$^4$ cm$^{-3}$) and low average ambient radiation fields over full galaxy scales \cite[(Cormier et al. 2015)]{cormier15}. Note, in contrast, the FIR line deficit seen in the most luminous sources \cite[(e.g. Luhman et al. 2003; Gracia-Carpio et al. 2011; Diaz Santos et al. 2013; Farrah et al. 2013)]{luhman03, gracia_carpio11,diaz_santos13,farrah13} where dustier HII regions are present leading to a high ionization parameter on global scales. On the other hand, the 88 \mic\ [OIII] line, requiring an ionization energy of 35 eV, is the brightest FIR cooling line in almost all of the dwarf galaxies \cite[(Cormier et al. 2015)]{cormier15} - not the [CII] line, as in the normal-metallicity galaxies. While higher stellar effective temperatures in the low metallicity star-bursting dwarf galaxies already can account for elevated [OIII] line emission compared to more metal-rich stellar atmospheres \cite[(e.g. Hunter et al. 2001)]{hunter01}, the predominance of the [OIII] line demonstrates the ease at which such hard photons can leak out of the medium surrounding stars and traverse the ISM on full galaxy scales (see also \cite[Cormier et al. 2012; Lebouteiller et al. 2012]{cormier12,lebouteiller12}).

The dwarf galaxies also show extreme [CII]/CO ratios (Fig.~\ref{CII_OI_CO}, {\it right}): while most normal star-forming galaxies are observed to have [CII]/CO $\sim$ 1\,000 to 4\,000, the low-metallicity galaxies can reach much higher ratios - up to an order of magnitude higher, or more, in [CII]/CO over galaxy-wide scales \cite[(e.g. Madden 2000; Cormier et al. 2014)]{madden00, cormier14}. Elevated \ciifir\ and \ciico\ values as seen above, seem to be attributed to the effects that are unique to the low-metallicity and clumpy ISM: a consequence of the lower dust abundance and a longer mean-free path of the FUV photons which penetrate molecular clouds, photodissociating the CO and leaving a relatively larger [CII]-emitting envelope surrounding a small CO core. This is consistent with the fact that we see such luminous 88 \mic\ [OIII] line emission over full galaxy-wide scales, and rarely can detect significant CO line emission. The decreased attenuation and the presence of intense, hard radiation fields, together facilitate the photodissociation of the molecular clouds, highlighting the different nature and structure of the ISM of low-metallicity galaxies.

  \section {H$_{2}$ mass in dwarf galaxies and the CO-dark gas}
\label{codark}

While CO is routinely used to trace the H$_{2}$ reservoir in local as well as high-z galaxies, its utility can be called into question for dwarf galaxies, which are notoriously deficient in CO \cite[(e.g. Schruba et al. 2012; Cormier et al. 2014 and references within)]{schruba12, cormier14}, leaving us with large uncertainties in quantifying the molecular gas mass, and compounding the difficulty in understanding the process of both local and global star formation in low-metallicity environments. 
 This road block is necessary to overcome in order to refine recipes for converting gas into stars under early-universe conditions and thus further our understanding of how galaxies evolve. We are left with the question: where is the bulk of the fuel for the obvious star formation taking place in many of the dwarf galaxies? 

While extreme observed [CII]/CO values of the dwarf galaxies suggest very active star formation, many of the dwarf galaxies are thought to be HI-rich rather than H$_{2}$-rich. The fact that their molecular gas reservoirs remain uncertain renders their location on the Schmidt-Kennicutt diagram ambiguous. Depending on how the total gas surface densities are determined (from HI, HI+H$_{2}$, CO, dust, or HI+H$_{2}$-modeled) several star-forming dwarfs deviate from the Schmidt-Kennicutt correlation with elevated star formation rates compared to their gas masses (Fig.~\ref{codark}; \cite[Cormier et al. 2014]{cormier14}).  

The lower dust abundance in dwarf galaxies allows the CO molecular clouds to be subjected more easily to photodissociation by UV photons, making the detections of the smaller CO cores with single-dish telescopes difficult. H$_{2}$, on the other hand, is photodissociated by absorption of Lyman-Werner band photons. These bands become optically thick at modest A$_{V}$ values and the H$_{2}$ becomes self-shielded. This self-shielding effect can leave a potentially significant reservoir of \hmol\ in the [CI] and [CII]-emitting regions, outside of the CO core \cite[(e.g. Roellig et al. 2006; Wolfire et al. 2010; Glover et al. 2012)]{roellig06, wolfire10, glover12}. This is the CO-dark molecular gas and the [CII], [CI] and/or other FIR fine-structure lines, may have the capability to trace this ``missing'' molecular gas that is not probed by CO. 
 This dark gas reservoir was first uncovered in a few low-metallicity galaxies using [CII] \cite[(Poglitsch et al. 1995; Israel et al. 1996; Madden et al. 1997)]{poglitsch95, israel96, madden97}, but \her\ has now increased the sample with deeper sensitivity. More recently [CII] observations of the Milky Way have found that on average the fraction of CO-dark gas, compared to the total \hmol\ can be as high as $\sim$ 30\% \cite[(Pineda et al. 2013)]{pineda13}. 
  \begin{figure}[!ht]
\begin{center}
 \includegraphics[width=2.6in]{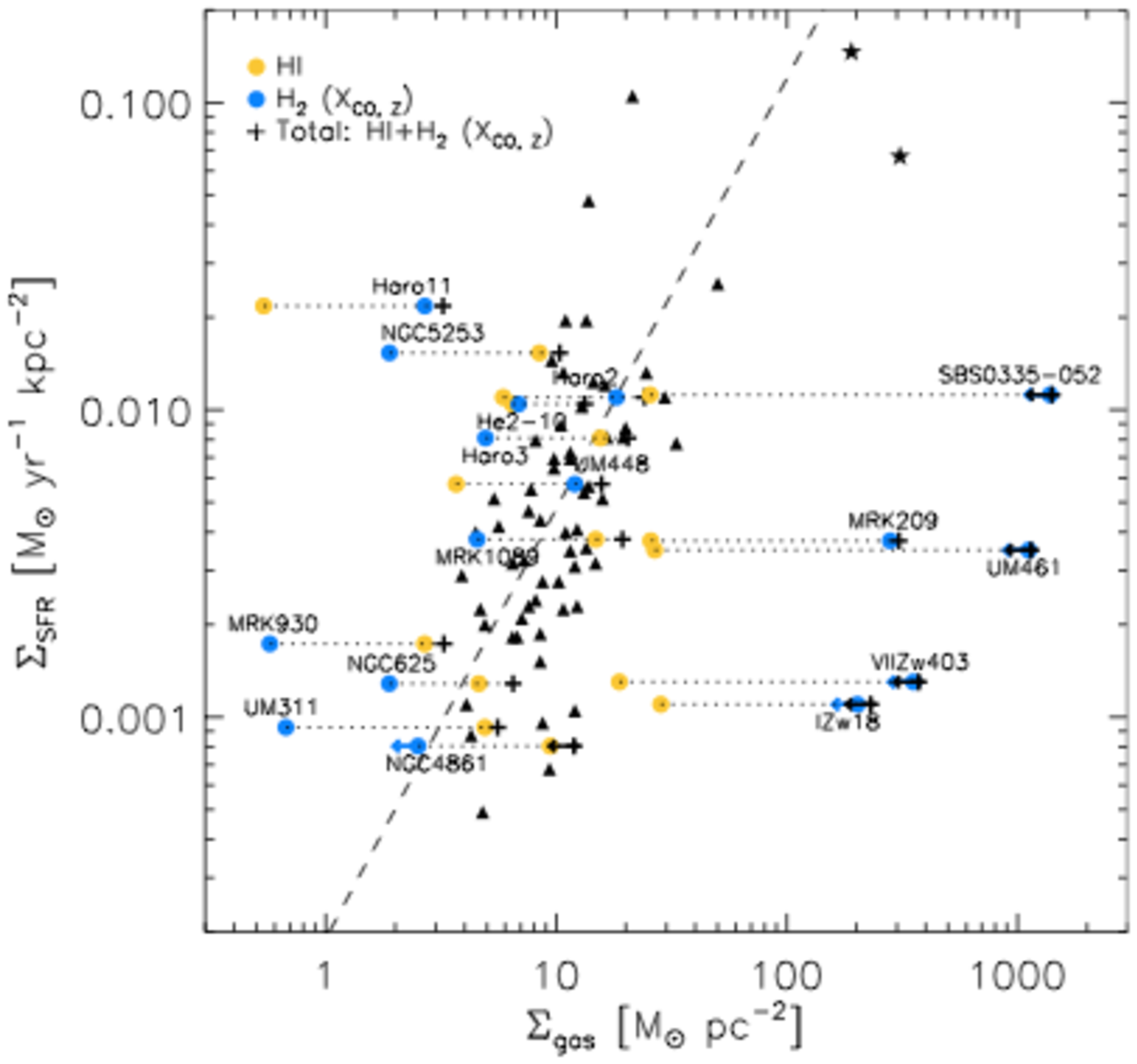}
\includegraphics[width=2.6in]{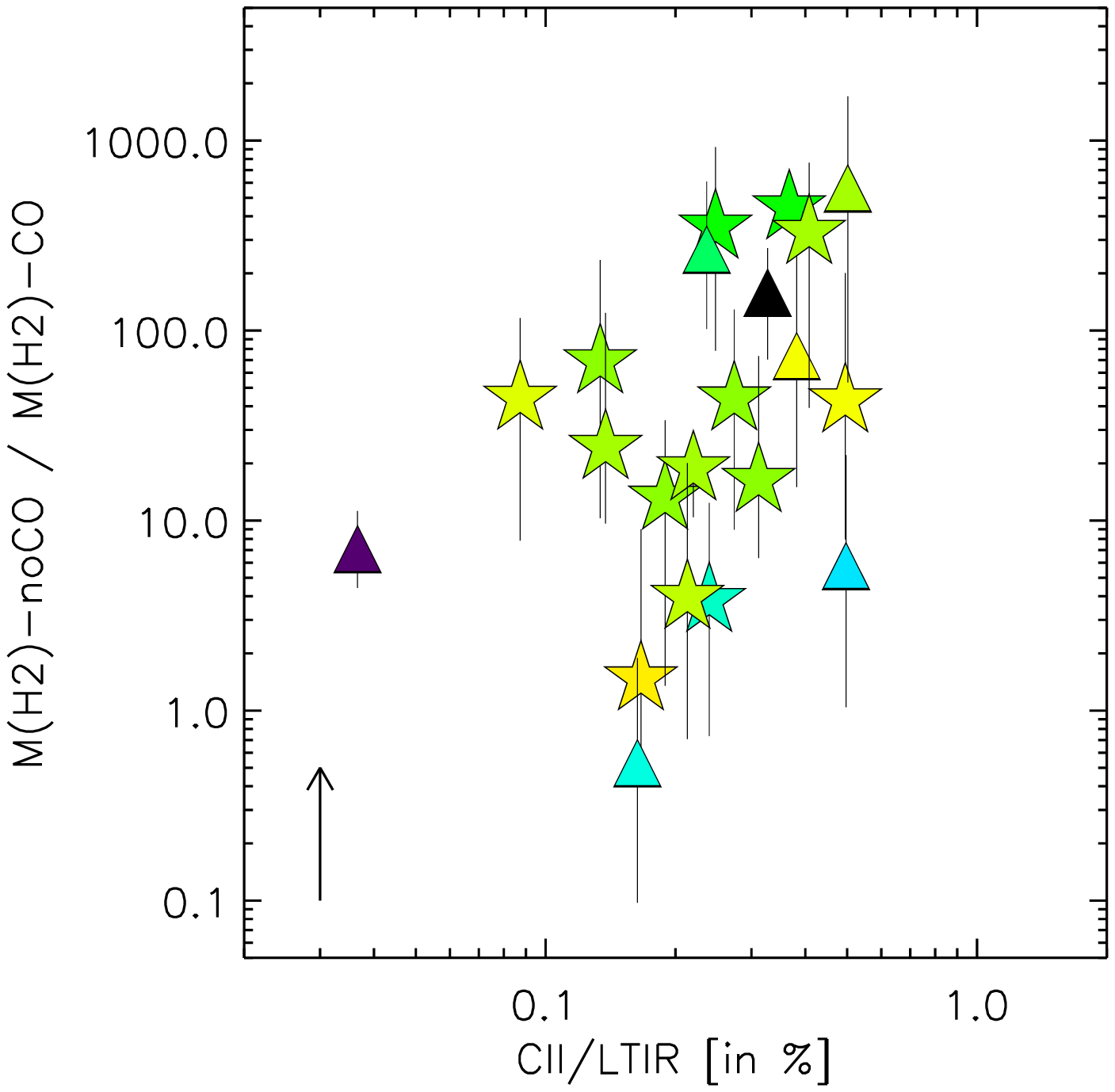}
\caption{\footnotesize{ {\it (left)} Star formation rate surface density ($\Sigma_{SFR}$) versus gas surface density ($\Sigma_{gas}$) for metal-rich starbursts and spirals from \cite[Kennicutt 1998]{kennicutt98} (black triangles) and the low metallicity DGS galaxies. Taking into account the large reservoir of HI brings the dwarf galaxies closer to the Schmidt-Kennicutt relationship (dotted line with power index of 1.4), while some remain as outliers to the right and some to the left of the relationship (see \cite[Cormier et al. 2014]{cormier14} for more details). {\it (right)} The CO-dark H$_{2}$ gas mass /H$_{2}$ determined from CO alone versus the observed [CII]/L$_{TIR}$ from modeling of the DGS (Madden et al. in prep). The triangles are upper limits due to lack of CO detections. The color indicates metallicity values as in the color scale in Fig.~3.}}
\label{codark}
\end{center}
\end{figure}

Determination of their true source of the molecular gas reservoir requires more than CO observations, which have proven to not be a reliable tracer of the total H$_{2}$ gas reservoir. The MIR and FIR fine-structure lines have recently been modeled for each galaxy of the DGS survey to quantify the predicted total H$_{2}$ reservoir determined from the best-fitted model (Cormier et al. in prep). This is then compared to the H$_{2}$-determined from the CO observations, the difference being the CO-dark H$_{2}$ (Fig.~\ref{codark}, {\it right}). The mass fraction of CO-dark gas / H$_{2}$ traced by CO, spans at least 4 orders of magnitude. The contribution of the CO-dark gas is greatest under conditions of the lowest A$_{V}$ environments as well low values of density and radiation field which correlate with higher [CII]/L$_{FIR}$ rates (with large scatter). Calibrating this empirically is work in progress. While we find that both [CII] and [CI] do indeed trace the CO-dark gas under the conditions we find for the dwarf galaxies, we also find that [CI] traces more of the CO-dark mass than [CII] \cite[(Madden et al. in prep)], making observations of the submm [CI] lines potentially the most valuable tracer of the CO-dark gas (see also \cite[Glover et al. 2015]{glover15}).

\begin{figure}[!ht]
\includegraphics[width=2.1in]{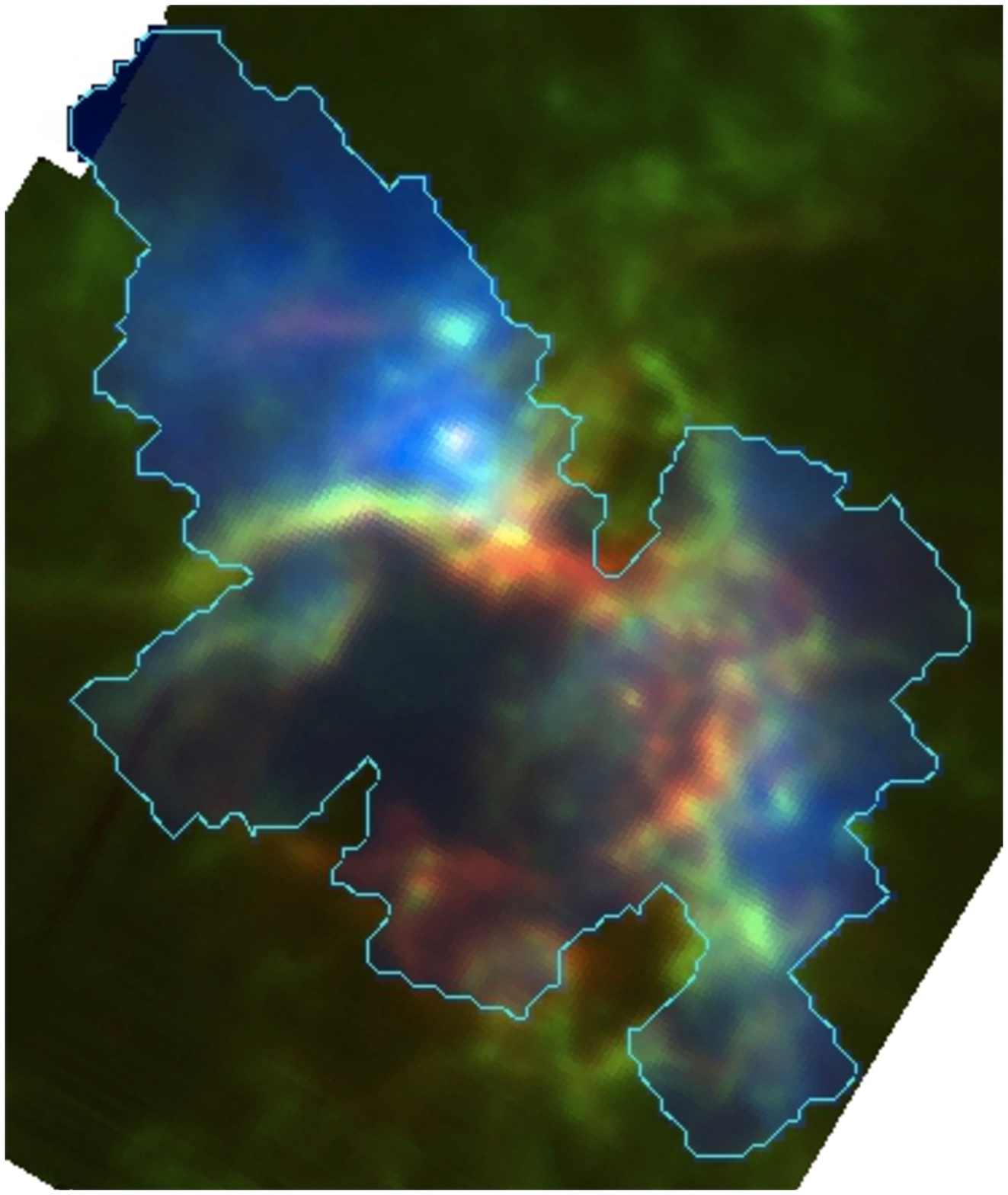}
\includegraphics[width=3.0in]{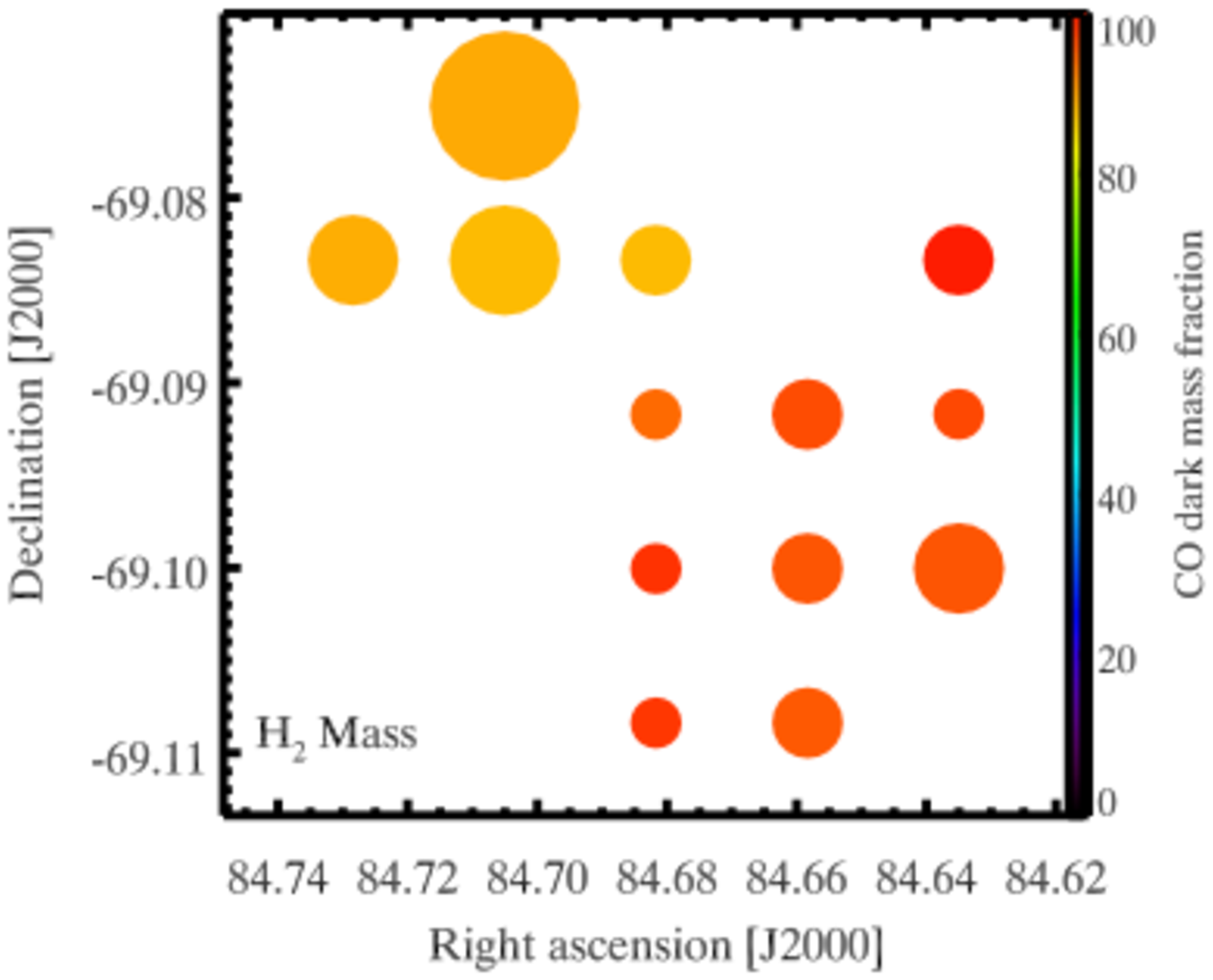}
\caption{\footnotesize{ {\it (left)} 3-color image of 30\,Doradus in the Large Magellanic Cloud. Red: [SIV]10.5 \mic; Green: [NeII]12.8 \mic;  Blue: PACS [CII]158 \mic. The blue contours outline the limit of the [CII] map. The different layers of the gas from the highly ionized medium near the star cluster R136 (image center with dark hole), to the PDR regions (blue), are evident (Chevance et al. 2015, submitted). {\it (right)} Fraction of the CO-dark gas calculated by comparing the predicted H$_{2}$ column density from PDR modeling to the H$_{2}$ derived from CO(1-0) observations. 
The color code is the CO-dark fraction which is $>$ 80\% throughout, with red the largest fraction of CO-dark gas. The size of the circles indicates the total H$_{2}$ mass, between 50 and 500 M$_{\odot}$ pc$^{-2}$.}}
\label{dor}
\end{figure}

To obtain greater detailed comprehension of the photodissociation of molecular clouds in the vicinity of a massive star cluster within a low-metallicity galaxy, Chevance et al. (2015, submitted) have studied the spatial variation of the ISM properties surrounding the R136 cluster in our nearest low metallicity galaxy, the Large Magellanic Cloud. Modeling the suite of MIR and FIR Herschel and Spitzer fine-structure emission lines (Fig.~\ref{dor},~{\it left}) has uncovered a significant dark-gas reservoir, where more than 80\% of the total H$_{2}$ is not traced by CO in the highly porous environment of 30\,Doradus (Fig.~\ref{dor},~{\it right}), demonstrating the value of the [CII] and [CI] lines.

\vskip -10mm
\section{Way forward: modeling complexities of the low-metallicity ISM} 

With the wealth of tracers of the galactic phases present today, it is the challenge to determine the local physical conditions within galaxies and to relate them to the global properties. With the wide range of MIR and FIR cooling lines available from \spitz\ and \her, for example, we are now able to model the physical conditions of the various phases of the gas and dust over full galaxy scales {\it as well as} spatially within galaxies. In this way we can obtain a more accurate prescription of the role of metallicity and star formation activity in governing the evolution of galaxies. 

Using limited MIR and FIR gas diagnostics to deduce the conditions for star formation and the structure of the galaxy is effectively averaging ensembles of PDRs and physically different gas phases mixed within a telescope beam. The challenge in moving forward with interpretation is to cover a range of diagnostics having varying critical densities and ionization potentials, to be able to characterize the dense and diffuse ionized gas and neutral atomic and molecular gas which have very different filling factors even for the range of dwarf galaxies \her\ has surveyed. Analysis of these various phases comprising galaxies cannot be done without considering the gas and dust tracers together. An example of one of the most thorough analysis that has been approached, given the wide selection of diagnostics for any one galaxy, is demonstrated in \cite[Cormier et al. (2012)]{cormier12} with $\sim$20 FIR fine-structure lines and CO observations as well as the photometry constraints to model the full SED of Haro\,11. While this galaxy was not resolved with the observations, the self-consistent photoionization and photodissociation modeling was able to determine the mass fraction and filling factor of the different ISM phases. This is the approach that is necessary now with the existence of many observational constraints. To go beyond this relatively simple, yet novel multiphase model though, including more realistic geometries, will really make that significant leap forward.

\section{Conclusion}

From studies of the MIR and FIR gas properties and MIR to submm dust properties, a picture of the structure of low-metallicity galaxies is emerging.
The decreased attenuation and the presence of intense, hard radiation fields conspire together to highlight the unique place the low-metallicity star-forming galaxies take compared to their more metal-rich counterparts.
The porosity of the ISM linked with the star formation activity in dwarf galaxies, accounts for the extreme [OIII]/[CII] and [CII]/CO. Much of the ISM is filled with hard photons which, due to the decrease in global dust attention, allows the permeating hard photons to heat the dust to higher temperatures, and fill a relatively large volume of the galaxy and successfully photodissociate CO clouds effectively, making single-dish CO observations challenging. We find that a significant reservoir of H$_{2}$, not traced by CO (the CO-dark gas), can exist and is traceable by [CII] and [CI]. 
  
\section{Acknowledgements}
This research was made possible through the financial support of the Agence Nationale de la Recherche (ANR) through the programme SYMPATICO (Program Blanc Projet NR-11-BS56-0023).


\begin{thebibliography}{}


\bibitem[Asano et al. (2013)]{asano13}
{{Asano},R.~S., {Takeuchi},T.~T., {Hirashita},H., \& {Inoue}, A.~K.} 2013,
\textit{Earth, Planets,, Space}, 65, 213
	
\bibitem[Cormier et al. (2012)]{cormier12}
{{Cormier}, D., {Lebouteiller}, V., {Madden}, S.~C., et al.} 2012,
\textit{A\&A}, 548, 20

\bibitem[Cormier et al. (2014)]{cormier14}
{{Cormier}, D., {Madden}, S.~C., {Lebouteiller}, V., et al.} 2014,
\textit{A\&A}, 564, 121

\bibitem[Cormier et al. (2015)]{cormier15} 
{{Cormier}, D., {Madden}, S.~C., {Lebouteiller}, V., et al.} 2015, 
\textit{A\&A}, 578, 53

\bibitem[Cortese et al. (2012)]{cortese12}
{{Cortese}, L., {Ciesla}, L., {Boselli}, A., et al.} 2012,
\textit{A\&A}, 540, 52

\bibitem[da Cunha et al. (2010)]{dacunha10}
{{da Cunha}, E., {Eminian}, C., {Charlot}, S., \& {Blaizot}, J.} 2010,
\textit{MNRAS}, 403, 1894

\bibitem[Dale et al. (2012)]{dale12}
{{Dale}, D.~A., {Aniano}, G., {Engelbracht}, C.~W., et al.} 2012,
\textit{ApJ}, 745, 95

\bibitem[D{\'{\i}}az-Santos et al. (2013)]{diaz_santos13}
{{D{\'{\i}}az-Santos}, T., {Armus}, L., {Charmandaris}, V., et al.} 2013,
\textit{ApJ}, 774, 68

\bibitem[Farrah et al. (2013)]{farrah13}
{{Farrah}, D., {Lebouteiller}, V., {Spoon}, H.~W.~W., et al.} 2013,
\textit{ApJ}, 776, 38

\bibitem[Galliano et al. (2011)]{galliano11}
{{Galliano}, F., {Hony}, S., {Bernard}, J.-P., et al.} 2011,
\textit{A\&A}, 536, 88

\bibitem[Glover \& Clark (2012)]{glover12}
{{Glover}, S.~C.~O. \& {Clark}, P.~C.} 2012,
\textit{MNRAS}, 421, 9

\bibitem[Glover \& Clark (2015)]{glover15}
{{Glover}, S.~C.~O. \& {Clark}, P.~C.} 2015,
\textit{2015arXiv150901939G}

\bibitem[Graci{\'a}-Carpio et al. (2011)]{gracia_carpio11}
{{Graci{\'a}-Carpio}, J., {Sturm}, E., {Hailey-Dunsheath}, S., et al.} 2011,
\textit{ApJ}, 728, 7

\bibitem[Griffin et al. (2010)]{griffin10}
{{Griffin}, M.~J., {Abergel}, A., {Abreu}, A., et al.} 2010,
\textit{A\&A} (Letter), 518, 3

\bibitem[Hunter et al. (2001)]{hunter01}
{{Hunter}, D.~A., {Kaufman}, M., {Hollenbach}, D.~J., et al.} 2001,
\textit{ApJ}, 553, 121

\bibitem[Israel et al. (1996)]{israel96}
{{Israel}, F.~P., {Maloney}, P.~R., {Geis}, N., et al.} 1996,
\textit{ApJ}, 465, 738

\bibitem[Kaufman, Wolfire \& Hollenbach (2006)]{kaufman06}
{{Kaufman}, M.~J., {Wolfire}, M.~G., \& {Hollenbach}, D.~J.} 2006,
\textit{ApJ}, 644, 283

\bibitem[Kennicutt (1998)]{kennicutt98}
{{Kennicutt}, Jr., R.~C.} 1998,
\textit{ApJ}, 498, 541
 
\bibitem[Lebouteiller et al. (2012)]{lebouteiller12}
{{Lebouteiller}, V., {Cormier}, D., {Madden}, S.~C., et al.} 2012,
\textit{A\&A}, 548, 91

\bibitem[Le Petit et al. (2006)]{lepetit06}
{{Le Petit}, F., {Nehm{\'e}}, C., {Le Bourlot}, J., \& {Roueff}, E.} 2006,
\textit{ApJS}, 164, 506

\bibitem[Luhman et al. (2003)]{luhman03}
{{Luhman}, M.~L., {Satyapal}, S., {Fischer}, J., et al.} 2003,
\textit{ApJ}, 594, 758

\bibitem[Madden et al. (1997)]{madden97}
{{Madden}, S.~C., {Poglitsch}, A., {Geis}, N., {Stacey}, G.~J., \& {Townes}, C.~H.} 1997,
\textit{ApJ}, 483, 200

\bibitem[Madden (2000)]{madden00}
{{Madden}, S.~C.} 2000,
\textit{NewAR}, 44, 249

\bibitem[Madden et al. (2013)]{madden13}
{{Madden}, S.~C., {R{\'e}my-Ruyer}, A., {Galametz}, M., et al.}, 2013,
\textit{PASP}, 125, 600

\bibitem[Pineda et al. (2013)]{pineda13}
{{Pineda}, J.~L., {Langer}, W.~D., {Velusamy}, T., \& {Goldsmith}, P.~F.} 2013,
\textit{A\&A}, 554, 103

\bibitem[Poglitsch et al. (1995)]{poglitsch95}
{{Poglitsch}, A., {Krabbe}, A., {Madden}, S.~C., et al.} 1995,
\textit{ApJ}, 454, 293

\bibitem[Poglitsch et al. (2010)]{poglitsch10}
{{Poglitsch}, A., {Waelkens}, C., {Geis}, N., et al.} 2010,
\textit{A\&A} (Letter), 518, 2

\bibitem[R{\'e}my-Ruyer et al. (2013)]{remy13}
{{R{\'e}my-Ruyer}, A., {Madden}, S.~C., {Galliano}, F., et al.} 2013,
\textit{A\&A}, 557, 95

\bibitem[R{\'e}my-Ruyer et al. (2014)]{remy14}
{{R{\'e}my-Ruyer}, A., {Madden}, S.~C., {Galliano}, F., et al.} 2014,
\textit{A\&A}, 563, 31

\bibitem[R{\'e}my-Ruyer et al. (2015)]{remy15}
{{R{\'e}my-Ruyer}, A., {Madden}, S.~C., {Galliano}, F., et al.} 2015,
\textit{A\&A}, 582, 121

\bibitem[R{\"o}llig et al. (2006)]{roellig06}
{{R{\"o}llig}, M., {Ossenkopf}, V., {Jeyakumar}, S., 
	{Stutzki}, J., \& {Sternberg}, A.} 2006,
\textit{A\&A}, 451, 917

\bibitem[Schruba et al. (2012)]{schruba12}
{{Schruba}, A., {Leroy}, A.~K., {Walter}, F., et al.} 2012,
\textit{AJ}, 143, 138

\bibitem[Skibba et al. (2011)]{skibba11}
{{Skibba}, R.~A., {Engelbracht}, C.~W., {Dale}, D., et al.} 2011,
\textit{ApJ}, 738, 89

\bibitem[Stacey et al. (1991)]{stacey91}
{{Stacey}, G.~J., {Geis}, N., {Genzel}, R., et al.} 1991,
\textit{ApJ}, 373, 423

\bibitem[Wolfire, Tielens \& Hollenbach (1990)]{wolfire90}
{{Wolfire}, M.~G., {Tielens}, A.~G.~G.~M., \& {Hollenbach}, D.} 1990,
\textit{ApJ}, 358, 116

\bibitem[Wolfire, Hollenbach \& McKee (2010)]{wolfire10}
{{Wolfire}, M.~G., {Hollenbach}, D., \& {McKee}, C.~F.} 2010,
\textit{ApJ}, 716, 1191

 \bibitem[Zhukovska (2015)]{zhukovska14}
{{Zhukovska}, S.} 2014, 
\textit{A\&A}, 562, 76

\end{thebibliography}
 \end{document}